\documentclass[twocolumn,amssymb,showpacs, amsmath,nobibnotes,nofootinbib,aps, prl]{revtex4}

\begin{document}

\title{Specific heat investigation of the magnetic ordering in two frustrated
spin-chain oxides: Ca$_{3}$Co$_{2}$O$_{6}$ and Ca$_{3}$CoRhO$_{6}$}
\author{V. Hardy$^{1,2}$, M. R. Lees$^{1}$, A. Maignan$^{2}$, S. H\'{e}bert$^{2}$,
D. Flahaut$^{2}$, C. Martin$^{2}$ and D. McK. Paul$^{1}$}
\affiliation{(1) Department of Physics, University of Warwick, CV4 7AL, Coventry, United\\
Kingdom\\
(2) Laboratoire CRISMAT, UMR 6508, Boulevard du Mar\'{e}chal Juin, 14050 Caen%
\\
Cedex,\\
France}
\date{\today}

\begin{abstract}
Specific heat measurements were carried out on the closely related
spin-chain oxides Ca$_{3}$Co$_{2}$O$_{6}$ and Ca$_{3}$CoRhO$_{6}$. Both
compounds consist of Ising magnetic chains that are arranged on a triangular
lattice. The spin coupling along and between the chains are ferromagnetic
and antiferromagnetic, respectively. Geometrical frustration is expected
from the combination of these magnetic features. The present study reports
that the specific heat data of these compounds exhibit similarities in the
spin-freezing process at low $T$, whereas striking differences exist in the
antiferromagnetic interchain ordering at higher $T$. These results are
discussed in connection with previous magnetization and neutron diffraction
data.
\end{abstract}
\maketitle

\section{Introduction}

\noindent Low dimensionality and frustration in magnetism continue to
attract a lot of interest, since both features are known to yield exotic
properties. Compounds of the family A'$_{3}$ABO$_{6}$ (where A' is Ca or Sr,
while A and B are transition metal elements) contain magnetic chains that
are arranged on a triangular lattice.\cite{ST01} Some of these compounds
possess an Ising-like character and an antiferromagnetic interchain
coupling, giving rise to geometrical frustration. The two such compounds
that have received most attention in the recent years are Ca$_{3}$Co$_{2}$O$%
_{6}$\cite{AA97,KA97a,KA97b,MA00} and Ca$_{3}$CoRhO$_{6}$.\cite
{NI99,NI01a,NI01b,NI02,SA02} It is worth emphasizing that the intrachain
coupling in these compounds is ferromagnetic, in contrast to the case of the
triangular Ising spin-chain antiferromagnets CsCoCl$_{3}$ and CsCoBr$_{3}$,
which have been intensively studied since the seventies.\cite{CO97}

The members of the A'$_{3}$ABO$_{6}$ family mentioned above seem to be the
first examples of systems with Ising ferromagnetic chains that are
antiferromagnetically coupled on a triangular lattice. Complex magnetic
ordering can be expected in such a situation, bearing some similarities with
the classical problem of 2D Ising triangular antiferromagnets.\cite{WA50}
The next-nearest interactions are known to be able to lift, at least
partially, the ground state degeneracy related to geometrical frustration,
allowing some degree of magnetic ordering to occur.\cite{ME74} As in the
earlier studies on CsCoCl$_{3}$ and CsCoBr$_{3}$, the existence of a
Partially Disordered Antiferromagnetic (PDA)\cite{ME} state has been
reported for both Ca$_{3}$Co$_{2}$O$_{6}$\cite{KA97a} and Ca$_{3}$CoRhO$_{6}$%
\cite{NI01a,NI01b}. In the PDA state, two thirds of the chains consist of
antiferromagnetically coupled ferromagnetic chains, while the remaining
third remain incoherent (disordered chains with zero net magnetization).
Note that such a state is very similar to the IOP1 phase found in more
recent theoretical work using the three-dimensional generalized six-state
clock model.\cite{TO02}

The similarities between the magnetic properties of Ca$_{3}$Co$_{2}$O$_{6}$
and Ca$_{3}$CoRhO$_{6}$ have led to the suggestion that both compounds have
similar phase diagrams with the same succession of magnetic states under
zero field:\cite{KA97a,NI01a} A paramagnetic state without order between the
chains at high $T$, a PDA state at intermediate temperature, and a frozen
spin (FS) state at low $T$. Recently, the existence of a PDA state in the
case of Ca$_{3}$Co$_{2}$O$_{6}$ was questioned,\cite{MA00,NI01b} while it
was supported by neutron diffraction data for the case of Ca$_{3}$CoRhO$_{6}$%
.\cite{NI01b} Furthermore, the distinction made between the PDA state and
the FS state was questioned on the basis of magnetic relaxation
measurements. \cite{SA02} To date, there have been no reports of specific
heat studies on these compounds. The aim of the present paper is to
investigate by specific heat measurements the magnetic ordering process in Ca%
$_{3}$Co$_{2}$O$_{6}$ and Ca$_{3}$CoRhO$_{6}$, in order to gain a deeper
insight into the similarities and differences that exist between these two
materials.

\section{Experimental details}

Ceramic samples of Ca$_{3}$Co$_{2}$O$_{6}$ and Ca$_{3}$CoRhO$_{6}$ were
prepared by standard solid state reaction in air. Stoichiometric proportions
of CaO, CoO$_{1.38}$ and RhO$_{2}$ were intimately ground and heated at 900 $%
^{\circ }C$ for a few days with intermediate grindings. Then the powder was
pressed in the form of bars and heated for one week at high temperature.
This sintering temperature was equal to 1000 $^{\circ }C$ and 1100 $^{\circ
}C$ for Ca$_{3}$Co$_{2}$O$_{6}$ and Ca$_{3}$CoRhO$_{6}$, respectively. X-ray
analysis confirmed the purity of these samples and showed that both
compounds crystallize in the space group R$\overline{3}$c. The cell
parameters derived from x-ray powder diffraction at room temperature are $a=$
0.907 nm\ and $c=$ 1.038 nm\ in Ca$_{3}$Co$_{2}$O$_{6}$, while they are $%
a=0.919$ nm\ and $c=1.071$ nm\ in Ca$_{3}$CoRhO$_{6}$. These values are
consistent with previously published results for these compounds.\cite
{AA97,NI99}

In the A'$_{3}$ABO$_{6}$ family, the A\ cation is on a trigonal prismatic
site, while B is on an octahedral site (see Fig. 1). Owing to their
different Crystalline Electric Field (CEF) environments, the former site
favours high spin states, whereas the latter favours low spin states.\cite
{AA97} In Ca$_{3}$CoRhO$_{6}$, the Co atoms occupy the prismatic sites and
the Rh atoms the octahedral sites.\cite{NI99} The valence state of the
transition elements in these compounds is still subject to controversy. In Ca%
$_{3}$Co$_{2}$O$_{6}$, the most likely configuration involves Co$^{3+}$ on
both sites, giving rise to $S=2$ on the prismatic sites and $S=0$ on the
octahedral sites, owing to the CEF effects mentioned above.\cite
{AA97,KA97a,KA97b,MA00} In Ca$_{3}$CoRhO$_{6}$, the reciprocal
susceptibility at high $T$ suggests that one deals with Co$^{2+}$ and Rh$%
^{4+}$, leading to an alternation between $S=3/2$ and $S=1/2$.\cite{NI99}
Nevertheless, recent neutron diffraction results at low $T$, as well as
electronic band structure calculations, have reported that both transition
metal elements of Ca$_{3}$CoRhO$_{6}$ should be in a trivalent state, giving
rise to the same alternation of $S=0$ and $S=2$ along the chains, as for Ca$%
_{3}$Co$_{2}$O$_{6}$.\cite{NI01b,WH03} All the configurations discussed
above are consistent with the fact that the measured saturation
magnetization $M_{S}$ in both compounds is found to be close to 4 $\mu _{B}$
/ f.u..\cite{AA97,NI01a}

A ceramic sample of Ca$_{3}$CaPtO$_{6}$ was used as a non-magnetic reference
in the specific heat analysis. It has been prepared from a stoichiometric
mixture of CaO and PtO$_{2}$, that was pelleterized in the form of bars and
heated in a evacuated silica ampoule at 800$^{\circ }C$ for 12 h. In Ca$_{3}$%
CaPtO$_{6}$, Pt$^{4+}$ has a $3d^{6}$ configuration which is supposed to
yield a low spin state ($S=0$) on the octahedral site. Magnetic measurements
confirmed that less than 0.15\% of the platinum ions in this compound are
magnetic.

The specific heat measurements were carried out by a two-tau relaxation
method (PPMS, Quantum Design). In order to determine the magnetic
contribution to the specific heat ($C_{M}$) in each compound, their lattice
contributions were derived from the data of Ca$_{3}$CaPtO$_{6}$ by using an
appropriate mass correction. We followed the procedure described by Bouvier 
{\it et al}.,\cite{BO91} that is based on a rescaling of the Debye
temperature. This method is quantitatively correct only at low temperature
(below about 50 K in our case). At higher temperatures, one can still be
confident in the overall temperature dependence of the magnetic contribution
to the specific heat, but the uncertainty in the $C_{M}$ values increases
with $T$. The magnetization measurements were carried out by means of a
SQUID magnetometer (MPMS, Quantum Design).

\section{Results and discussion}

Figure 2 shows the susceptibility curves for Ca$_{3}$Co$_{2}$O$_{6}$ and Ca$%
_{3}$CoRhO$_{6}$ in an applied field of 0.1 T, that were recorded in the
Zero Field Cooled (ZFC) and Field Cooled Cooling (FCC) modes. The high
temperature regimes lead to positive Curie-Weiss temperatures indicative of
ferromagnetic intrachain coupling, in agreement with the literature.\cite
{AA97,NI99} These Curie-Weiss temperatures were found to be equal to 30 K 
\cite{AA97,KA97a} and 150 K \cite{NI99} in Ca$_{3}$Co$_{2}$O$_{6}$ and Ca$%
_{3}$CoRhO$_{6}$ respectively, which indicates that the intrachain coupling $%
J$ is significantly larger in the latter compound. Figure 2 shows that, for
both compounds, $\chi $ undergoes a pronounced upturn below a characteristic
temperature that is close to 25 K and 90 K, for Ca$_{3}$Co$_{2}$O$_{6}$ and
Ca$_{3}$CoRhO$_{6}$ respectively. Since long-range ordering does not exist
in one dimension,\cite{REV} such features cannot simply be attributed to
intrachain ferromagnetic transitions. They are connected to the interchain
(antiferromagnetic) ordering, in accordance with the emergence of magnetic
Bragg peaks found below $T_{N}=26$ K and $T_{N}=89.9$ K, in Ca$_{3}$Co$_{2}$O%
$_{6}$\cite{AA97} and Ca$_{3}$CoRhO$_{6}$,\cite{NI01a} respectively. This
shift in $T_{N}$ indicates that the interchain coupling $J^{\prime }$ is
also larger for Ca$_{3}$CoRhO$_{6}$ than for Ca$_{3}$Co$_{2}$O$_{6}$. A part
of the apparent difference in the shape of $\chi (T)$ between Ca$_{3}$Co$%
_{2} $O$_{6}$ and Ca$_{3}$CoRhO$_{6}$ is simply a matter of temperature
range. As shown in the insets of Fig. 2, the data for both compounds look
quite similar after rescaling by $T_{N}$. In particular, it should be noted
that the ZFC and FCC curves for Ca$_{3}$Co$_{2}$O$_{6}$ and Ca$_{3}$CoRhO$%
_{6}$ all flatten at very low temperatures. This low $T$ regime, that is
seen in both compounds, has been ascribed to the onset of a frozen spin
state \cite{KA97a} (more specifically a frozen PDA state\cite{NI01a} in the
case of Ca$_{3}$CoRhO$_{6}$).

Beyond these similarities, Fig. 2 also displays several differences between
the two compounds. First of all, the magnetic transitions take place at much
higher temperatures in Ca$_{3}$CoRhO$_{6}$ than in Ca$_{3}$Co$_{2}$O$_{6}$,
and the susceptibility values are significantly lower in the former case.
The rounding of $\chi (T)$ above $T_{N}$ is also found to be more pronounced
in Ca$_{3}$CoRhO$_{6}$ than in Ca$_{3}$Co$_{2}$O$_{6}$. It should be noted
that such a rounding probably originates from fluctuations related to the
antiferromagnetic interchain coupling. The second main difference concerns
the shape of the ZFC curve which displays a double peak structure in the
case of Ca$_{3}$Co$_{2}$O$_{6}$.\cite{rq0} This particular behaviour -which
is connected with a marked time dependence\cite{KA97b,MA00} of the
magnetization in this intermediate temperature range- will be discussed
elsewhere.

Figure 3 shows the $C(T)$ curves recorded under zero field in Ca$_{3}$Co$%
_{2} $O$_{6}$ and Ca$_{3}$CoRhO$_{6}$. This data set points to a fundamental
difference between the two compounds: While both compounds exhibit a sudden
increase of $\chi (T)$ accompanied by the appearance of Bragg peaks below a
characteristic temperature $T_{N}$ (about 26 K and 90 K in Ca$_{3}$Co$_{2}$O$%
_{6}$ and Ca$_{3}$CoRhO$_{6}$, respectively), {\it there is a specific heat
peak associated with this transition in} Ca$_{3}$Co$_{2}$O$_{6}$ {\it but
not in} Ca$_{3}$CoRhO$_{6}$. Owing to an antiferromagnetic coupling between
Ising spin-chains that are arranged on a triangular lattice, the magnetic
ordering is expected to be impeded by geometrical frustration effects in
these compounds. However, next-nearest-neighbours ($nnn$) interactions can
weaken such frustration effects, allowing the establishment of a long-range
ordering.\cite{TA95} In Ca$_{3}$CoRhO$_{6}$, a recent neutron diffraction
study claimed that the magnetic ordering below $T_{N}\simeq 90$ K was a true
PDA state.\cite{NI01b} This state is expected in the presence of a weak
ferromagnetic $nnn$ interaction.\cite{ME} It should be noted that a peak in $%
C(T)$ at the transition from the paramagnetic to the PDA state has never
been observed in\ CsCoCl$_{3}$, despite the fact that this compound is
supposed to be a prototypical example of a material exhibiting such a
transition.\cite{ME} It should be also emphasized, however, that Monte Carlo
simulations in the case of pure 2D Ising triangular antiferromagnets
demonstrated that the transitions to partially ordered states can have very
different signatures in the $C(T)$ data (including prominent peaks)
depending on the values of the $nnn$ interactions.\cite{TA95}

Since the partially ordered state below $T_{N}$ is still poorly understood
in the case of Ca$_{3}$Co$_{2}$O$_{6}$, one cannot rule out the possibility
that the existence of a peak in $C(T)$ for this compound and not for Ca$_{3}$%
CoRhO$_{6}$ results from a totally different nature of the ordering process
in these two compounds. Nevertheless, this would be surprising given their
structural similarity, and it seems more reasonable to expect that the
ordering process is qualitatively similar in both materials. In this case,
the absence of a peak in specific heat for Ca$_{3}$CoRhO$_{6}$ should simply
be ascribed to a more progressive character of the magnetic transition below 
$T_{N}$. We note that such a more continuous character of the transition in
Ca$_{3}$CoRhO$_{6}$ is consistent with the $\chi (T)$ curve that exhibits a
smoother upturn below $T_{N}$ as compared to that observed in Ca$_{3}$Co$_{2}
$O$_{6}$ (see Fig. 2).\cite{rq} Similarly, the increase in the amplitude of
the antiferromagnetic Bragg peak (100) below $T_{N}$ appears to be smoother
in Ca$_{3}$CoRhO$_{6}$ than in Ca$_{3}$Co$_{2}$O$_{6}$ (see Ref. \cite{NI01a}
and \cite{AA97}). In Ca$_{3}$Co$_{2}$O$_{6}$, the amplitude of this Bragg
peak reaches its maximum value well above half $T_{N}$,\cite{AA97} in
contrast with the case of Ca$_{3}$CoRhO$_{6}$.\cite{NI01a} At this stage,
the question remains, why the magnetic transition is more progressive in Ca$%
_{3}$CoRhO$_{6}$. In spite of their structural and electronic similarities,
we have already noted that the two compounds clearly differ by their values
of the intrachain coupling $J$, interchain coupling $J^{\prime }$, and
critical temperature $T_{N}$ (this last parameter being essentially
determined by the two previous ones). In a situation of geometrical
frustration, as in our case, it is worth noticing that the entropic effects,
which tend to impede the ordering process, should be enhanced in the case of
Ca$_{3}$CoRhO$_{6}$ because of the higher $T\;$range of the transition. This
could spread out so much the entropy change at $T_{N}$ that no peak can be
detected in the $C(T)$ curve. In other respects, the large variations in $J$
and $J^{\prime }$ that are observed between Ca$_{3}$Co$_{2}$O$_{6}$ and Ca$%
_{3}$CoRhO$_{6}$ suggest that the $nnn$ interactions may also be different
in these two compounds. According to Ref. 20, a difference between Ca$_{3}$Co%
$_{2}$O$_{6}$ and Ca$_{3}$CoRhO$_{6}$ about the values of these latter
parameters could play a role in the difference that is observed in the
signature of the ordering process in $C(T)$.

Let us now consider the insets of Fig. 3 that focus on the low temperature
regime. A spin-freezing has been reported to occur in both compounds at low $%
T$.\cite{KA97a,NI01a} In addition to the divergence between the ZFC and FC
curves, the spin-freezing leads to a large hysteresis in $M(H)$ curves. This
latter phenomenon was reported to take place below $\sim $8 K and $\sim $30
K in Ca$_{3}$Co$_{2}$O$_{6}$\cite{KA97b,MA00} and Ca$_{3}$CoRhO$_{6}$, \cite
{NI99} respectively. The $C/T$ vs. $T^{2}$ plot of Ca$_{3}$Co$_{2}$O$_{6}$
shows a clear change of regime below $\sim $8 K that can be associated with
this crossover to the FS state. Because of the low crossover temperature in
Ca$_{3}$Co$_{2}$O$_{6}$, this feature is directly visible in the total
specific heat, contrary to the case of \ Ca$_{3}$CoRhO$_{6}$ for which the
lattice contribution must be removed (see below). Furthermore, the $C/T$ vs. 
$T^{2}$ plots in the insets of Fig. 3 demonstrate the existence of a linear
term $\gamma T$ at low temperature, with approximately the same value $%
\gamma \simeq 10$ mJ K$^{-2}$ mol$^{-1}$ in both compounds. Such a linear
term in specific heat is consistent with the magnetic disorder expected in
these frozen spin (FS) states.

Figure 4 displays the magnetic specific heat of Ca$_{3}$Co$_{2}$O$_{6}$ and
Ca$_{3}$CoRhO$_{6}$ in zero field. Apart from the 3D ordering peak in Ca$_{3}
$Co$_{2}$O$_{6}$, both curves contain a broad maximum at intermediate
temperature. Even if the subtraction of the lattice contribution leads to
more uncertain values as $T$ is increased, this uncertainty will not remove
these features. Such broad maxima are probably related to the development of
short-range correlations along the chains, as theoretically expected in 1D
systems\cite{REV} and experimentally observed in various Ising spin-chain
compounds.\cite{EXP} Figure 4 also reveals that $C_{M}(T)$ for Ca$_{3}$CoRhO$%
_{6}$ exhibits a change in temperature dependence around 25 K. Below $\sim 20
$ K, $C_{M}(T)$ takes very small values, and then increases rapidly with
temperature above $\sim 30$ K (see inset of Fig. \ 4). This change of regime
correlates very well with the sudden increase in the ZFC curve around 25 K
which was attributed to the transition from a PDA to a F-PDA state [see Fig
2(b)]. In Ca$_{3}$Co$_{2}$O$_{6}$, there is also a quite good agreement
between the large increase in the ZFC curve at low $T$ (centered around 6 K)
and the change of regime in $C_{M}(T)$ around 8 K (see inset of Fig. \ 4).
The data for both compounds indicate that the crossover to the FS state has
a clear signature in the specific heat. On a simple level, one may expect
the freezing of the spin configuration to lead to a marked decrease in the
value of the specific heat. It must be noted that, for both compounds, the
change of regime associated with the crossover between the PDA and F-PDA
states is rather gradual.

The application of a magnetic field can profoundly affect the magnetic state
in Ca$_{3}$Co$_{2}$O$_{6}$ and Ca$_{3}$CoRhO$_{6}$. On the basis of
magnetization measurements,\cite{AA97,KA97a,KA97b,MA00,NI01a} a
ferrimagnetic phase has been reported to occur in both compounds under
intermediate fields (e.g. 2 T). In Ca$_{3}$Co$_{2}$O$_{6}$, the existence of
such a phase is well supported by the plateau found at one third of the
saturated magnetization in $M(H)$ curves.\cite{AA97,KA97a,KA97b,MA00} For Ca$%
_{3}$CoRhO$_{6}$, the signature of this ferrimagnetic transition in
magnetization measurements is less clear,\cite{NI01a} although a neutron
diffraction study claimed the existence of a ferrimagnetic state in a field
of 2 T.\cite{NI01b} Figure 5 displays enlargements of $C(T)$ around $T_{N}$
in 0 T and 2 T for both compounds. \cite{rq1} In Ca$_{3}$Co$_{2}$O$_{6}$,
one finds a peak in $C(T)$ in 2 T which is even more pronounced than the
peak seen in zero-field. This is consistent with the fact that the low $T$
phase of the transition is better ordered in 2 T (ferrimagnetism) than in 0
T (disordered antiferromagnetism). In contrast, for Ca$_{3}$CoRhO$_{6}$ the
curves in 0 T and 2 T are found to be perfectly superimposed on each other.
\cite{rq2} It must be emphasized that the absence of a peak in $C(T)$ in 2 T
for Ca$_{3}$CoRhO$_{6}$ calls into question the existence of a ferrimagnetic
state in this compound. At least, this result eliminates the possibility
that there is a simple paramagnetic-to-ferrimagnetic phase transition under
field, as is found in Ca$_{3}$Co$_{2}$O$_{6}$.

Although there is no peak in 2 T or 0 T in Ca$_{3}$CoRhO$_{6}$, one can
detect a change in the slope of $C(T)$ around $T_{N}$, as illustrated in
Fig. 5(b). This crossover takes place around 87 K. This value corresponds
well to the kink in $\chi (T)$, and so may be correlated with the beginning
of the magnetic ordering. A comparison of Fig. 5(a) and 5(b) clearly points
to a much smoother magnetic interchain ordering in Ca$_{3}$CoRhO$_{6}$ than
in Ca$_{3}$Co$_{2}$O$_{6}$, whatever the field value.

\section{Conclusion}

Specific heat measurements were carried out for the first time on Ca$_{3}$Co$%
_{2}$O$_{6}$ and Ca$_{3}$CoRhO$_{6}$, the two most studied members of a new
family of spin-chain oxides. Both compounds exhibit peculiar magnetic
properties that are related to a specific combination of features: Ising
chains arranged on a triangular lattice, with a ferromagnetic intrachain
coupling and an antiferromagnetic interchain coupling. Previous
magnetization and neutron diffraction studies have shown that the properties
of both compounds are similar in many points, as expected from their related
magnetic chain structure, but some differences have also been reported.\cite
{AA97,KA97a,KA97b,MA00,NI99,NI01a,NI01b,NI02,SA02}

The present specific heat study sheds new light on the comparison between
these two compounds: (i) Ca$_{3}$Co$_{2}$O$_{6}$ exhibits a peak in $C(T)$
at the beginning of the interchain ordering, while there is not such a peak
in the case of Ca$_{3}$CoRhO$_{6}$; (ii) The crossover to the frozen spin\
states are visible in the specific heat of both Ca$_{3}$Co$_{2}$O$_{6}$ and
Ca$_{3}$CoRhO$_{6}$, below about 8 K and 25 K, respectively. The specific
heat in these FS states is characterized by a linear term of amplitude close
to $10$ mJ K$^{-2}$ mol$^{-1}$; (iii) In contrast to the situation for Ca$%
_{3}$Co$_{2}$O$_{6}$, the specific heat data for Ca$_{3}$CoRhO$_{6}$ is not
consistent with a clear ferrimagnetic-to-paramagnetic transition around $%
T_{N}$ in intermediate fields (e.g. 2 T).

Despite their structural similarity, Ca$_{3}$Co$_{2}$O$_{6}$ and Ca$_{3}$%
CoRhO$_{6}$ have significantly different intrachain and interchain coupling
constants, which take larger values in the latter compound. The origin of
such variations between Ca$_{3}$Co$_{2}$O$_{6}$ and Ca$_{3}$CoRhO$_{6}$, as
well as their connection with the differences observed in the specific heat
and magnetization features of these compounds, remain open questions at the
present time.

\bigskip This work is supported by a EPSRC Fellowship grant (GR/R94299/01$)$
to one of the authors (V. H. ). The authors also acknowledge financial
support from the CNRS in the frame of a CNRS / Royal Society exchange
program (n$^{\circ }$ 13396).

\section{Figure captions}

Fig. 1: Schematic drawings of the structure of A'$_{3}$ABO$_{6}$-type
compounds. The dark and light polyhedra represent AO$_{6}$ trigonal prisms
and BO$_{6}$ octahedra, respectively. The shaded circles denote A' atoms.
(a) Perspective view showing the $\left[ \text{ABO}_{6}\right] _{\infty }$%
chains running along the hexagonal $c$-axis. (b) Projection along the
hexagonal $c$-axis. Solid lines emphasize the triangular arrangement of the
chains in the $ab$ plane.

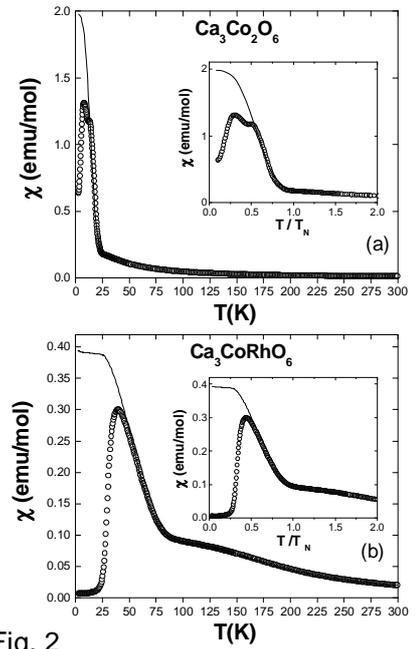
Fig. 2: Magnetic susceptibility under 0.1 T in (a) Ca$_{3}$Co$_{2}$O$_{6}$
and (b) Ca$_{3}$CoRhO$_{6}$. The circles and the solid lines correspond to
the zero-field-cooled and field-cooled cooling modes, respectively. The
insets show enlargements of these curves in reduced temperature scales, $%
T/T_{N}$, with $T_{N}$ equal to 26 K and 90 K in Ca$_{3}$Co$_{2}$O$_{6}$ and
Ca$_{3}$CoRhO$_{6}$, respectively.

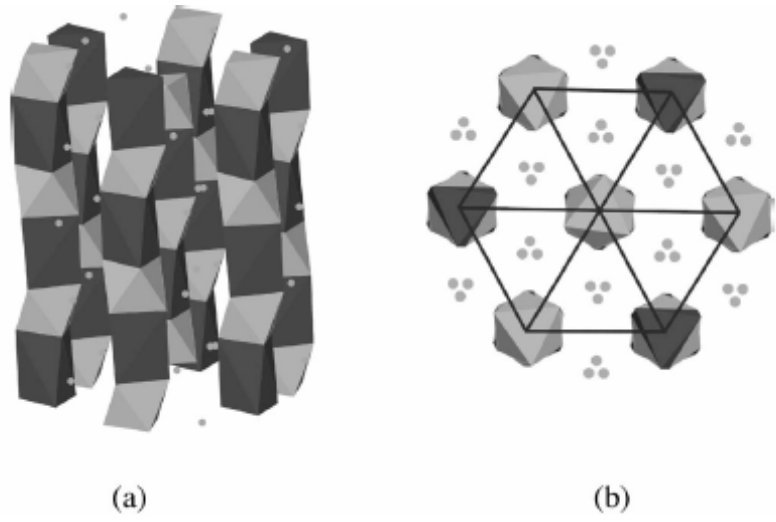
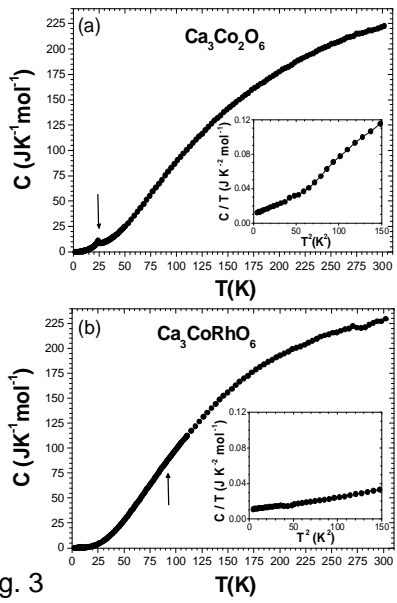
Fig. 3: Specific heat under zero field in (a) Ca$_{3}$Co$_{2}$O$_{6}$ and
(b) Ca$_{3}$CoRhO$_{6}$. The arrows denote the temperatures of the upturns
in $\chi (T)$. The insets display the very low $T$ regimes $(T<12$ K) in $%
C/T $ vs. $T^{2}$ plots.

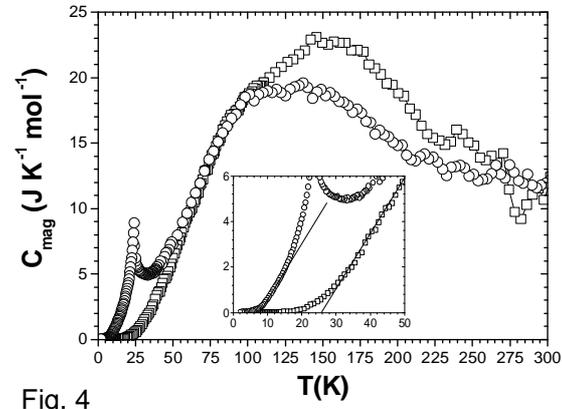
Fig. 4: Magnetic contribution to specific heat under zero field, in Ca$_{3}$%
Co$_{2}$O$_{6}$ (circles) and Ca$_{3}$CoRhO$_{6}$ (squares). The inset is an
enlargement of the low $T$ range, with straight lines emphasizing the
crossovers found around 8 K and 25 K in Ca$_{3}$Co$_{2}$O$_{6}$ and Ca$_{3}$%
CoRhO$_{6}$, respectively.

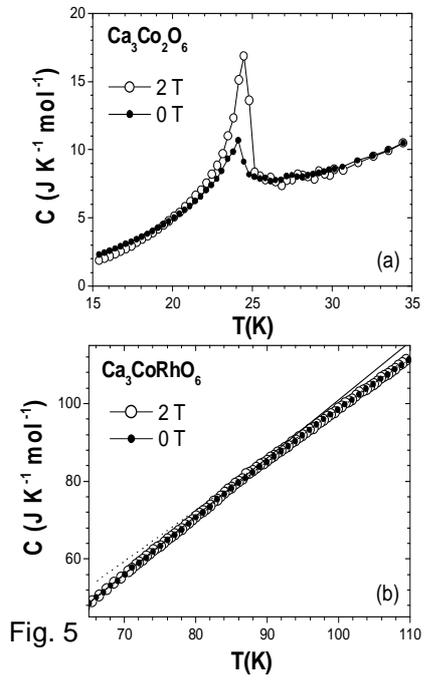
Fig. 5: Temperature dependence of the specific heat of (a) Ca$_{3}$Co$_{2}$O$%
_{6}$ and (b) Ca$_{3}$CoRhO$_{6}$, around the interchain magnetic ordering ($%
T_{N}$), under 0 T\ and 2 T (solid and empty circles, respectively). The
solid and dashed lines in (b) are linear fittings to $C(T)$ below and above
87 K, respectively.


\begin{thebibliography}{99}
\bibitem{rq0}  Such a feature can also be observed on the susceptibility
curves of references 2, 3 and 5.
\bibitem{rq}  Note that this difference remains perceptible in the insets of
Fig. 2 with reduced temperature scales.
\bibitem{rq1}  No noticeable difference could be detected between ZFC and FC
measurements under 2 T.
\end{thebibliography}
\end{document}